\documentstyle[aps,prl]{revtex}
\begin{document}
\draft
\twocolumn[\hsize\textwidth\columnwidth\hsize\csname @twocolumnfalse\endcsname
\title{Quantum spin ladder with dynamic phonons.
Bound states of magnons and phonons.}
\author{O.P. Sushkov}
\address{School of Physics, University of New South Wales, Sydney 2052, Australia}

\maketitle

\begin{abstract}
We consider the two-chain Heisenberg ladder
 with antiferromagnetic interactions.
Our approach is based on the description of spin excitations
as triplets above a strong-coupling singlet ground state.
It is shown that interaction with dynamic phonons drastically changes the
dispersion of the elementary triplet excitation (magnon) and the spin 
structure factor.
It is also shown that a new triplet excitation appears in the
spectrum: the bound state of a magnon and a phonon.
\end{abstract}

]

Quantum spin ladders  
have recently become   a subject of considerable interest,  mainly because 
of their relevance to a variety of quasi one-dimensional materials 
\cite{Dagotto}. It is well established that the two-leg $S=1/2$ 
spin  ladder has a gapped excitation spectrum \cite{Copalan,Exact,DMRG,O}.
Quite recently  a  state with the energy  above the one-particle excitation 
level but below the two-particle continuum was observed in 
(VO)$_{2}$P$_{2}$O$_{7}$ 
which can be described as a dimerized chain \cite{Garrett}.
This result triggered theoretical work on two-particle bound states
 in this model \cite{BS}. The above results raised the important issue of 
possible low-lying two-particle excitations  in quasi one-dimensional models
with a gapped spectrum.  It has been shown that in a two-leg ladder
bound states with total spin S=0 and total spin S=1 always exist below
the continuum \cite{Dam,us1}. Similar bound states exist also in many
two-dimensional quantum spin systems \cite{us1,us2}. To complete the overview
of pure spin systems (no dynamic phonons) we should mention that
for spin chains with small explicit dimerization the spectrum 
can be even more complex consisting of multiple triplet and singlet states
\cite{Affleck,Uhrig}.

The importance of dynamic phonons was realized a long time ago in relation to
the spin-Peirls transition \cite{Cross}. In the paper \cite{Kh} a simple model 
Hamiltonian for description of the combined spin-phonon dynamics in a spin 
chain has been suggested. Numerical study of this Hamiltonian has 
shown that many physical properties differ substantially from 
those derived in the static limit \cite{Wel}. This important observation 
raises the question of the role of phonons in the dynamics of the spin ladder.
In the present work, using the analytical approach developed in \cite{K},
we demonstrate that dynamic phonons qualitatively change the spectrum of
triplet excitations in the two-leg spin ladder: 1) an additional triplet 
appears as a magnon-phonon bound state, 2) the  magnon
dispersion attains discontinuity. 3)q-dependence of the spin structure 
factor is substantially different from that without phonons.

Following the idea of work \cite{Kh} we consider
the Hamiltonian of  two coupled $S=1/2$ chains (spin ladder)
with local phonons coupled to the rung spin-spin interaction:
\begin{eqnarray}
\label{H}
H&=&\sum_i\left\{J{\bf S}_i.{\bf S}_{i+1}+
J{\bf S}_i^{\prime}.{\bf S}_{i+1}^{\prime}
+J_{\perp}{\bf S}_i.{\bf S}_i^{\prime} \right\}\\
&+&\sum_i\left\{g(b_i^{\dag}+b_i){\bf S}_i.{\bf S}_i^{\prime} 
+\omega_0 b_i^{\dag} b_i  \right\}.\nonumber
\end{eqnarray}
Here $b_i^{\dag}$ and $b_i$ are the phonon creation and annihilation
operators. We assume the optical phonon to be dispersionless.
The  intra-chain ($J$) and the inter-chain ($J_{\perp}$) 
interactions are assumed 
antiferromagnetic $J,J_{\perp} >0$. 

Following works \cite{Copalan,K} we consider the set of the rung 
singlets as the unperturbed ground state, introduce a triplet creation 
operator on the rung $t_{\alpha i}^{\dag}$, $\alpha=x,y,z$, and hence map the 
Hamiltonian (\ref{H}) to
\begin{eqnarray}
\label{H1}
H&=&\sum_{i,\alpha,\beta}\left\{J_{\perp}t_{\alpha i}^{\dag}t_{\alpha i}+
{J\over{2}}\left(t_{\alpha i}^{\dag}t_{\alpha i+1}+
t_{\alpha i}^{\dag}t_{\alpha i+1}^{\dag}+\mbox{h.c.}\right) \right.
\nonumber \\
&+&\left. {J\over{2}}\left(t_{\alpha i}^{\dag}
t_{\beta i+1}^{\dag}t_{\beta i}t_{\alpha i+1}
-t_{\alpha i}^{\dag}t_{\alpha i+1}^{\dag}t_{\beta i}t_{\beta i+1}\right)
\right\}+\\
&U& \sum_{i,\alpha \beta}
t_{\alpha i}^{\dagger}t_{\beta i}^{\dagger}t_{\beta i}t_{\alpha i}
+\sum_{i,\alpha}\left\{g(b_i^{\dag}+b_i) 
t_{\alpha i}^{\dag}t_{\alpha i}
+\omega_0 b_i^{\dag} b_i  \right\},\nonumber
\end{eqnarray}
where $U \to \infty$ is a fictitious interaction which enforces no double
occupancy constraint $t_{\alpha i}^{\dag}t_{\beta i}^{\dag}=0$.
For shortness we call the excitation corresponding to $t^{\dag}$ 
(``elementary'' triplet) a magnon.

We will treat the interaction with phonons by perturbation theory and therefore
at the first step we have to diagonalize the Hamiltonian (\ref{H1}) at
$g=0$. The technique to perform this diagonalization was suggested
in Ref. \cite{K} with further improvement (rainbow diagrams and attraction 
in the singlet channel) developed in relation to the analysis of the
two-dimensional $J_1-J_2$ model \cite{us2}. Applying these techniques we
find that the normal Green's function of the magnon is of the
form
\begin{equation}
\label{GN}
G(k,\epsilon) = \frac{u_k^{2}}{\epsilon - \Omega_k+i\delta} + 
\frac{v_k^{2}}{-\epsilon - \Omega_k+i\delta} + G_{inc}. 
\end{equation}
Coupled Dyson's equations for the effective Bogoliubov's parameters
$u_k$ and $v_k$ and for the magnon dispersion $\Omega_k$ are derived in 
Refs.\cite{K,us2}. In the limit $J_{\perp}\gg J$ the solution is very simple: 
$u_k\approx1$,  $v_k\approx -(J/2J_{\perp})\cos k$, and
\begin{equation}
\label{e}
\Omega_k\approx J_{\perp}+J\cos k.
\end{equation}
The magnon dispersion at $J_{\perp}=J$ found as a result of numerical
solution of Dyson's equations  is presented in Fig.1 (long-dashed line).
For comparison we show by dots the results obtained from dimer series 
expansions \cite{O}.

Probing with neutrons is the experimental method commonly used to 
measure excitation spectra of magnetic materials.
The inelastic neutron scattering cross-section is directly proportional
to the dynamic structure factor:
$ S_{u}(k,\epsilon) = 
 \int e^{i\epsilon t} \langle S^{u}_{z}(k,t)S^{u}_{z}
(-k,0) \rangle dt$, where we assume that the
 transverse (along the rungs) momentum $k_{\perp}=\pi$, i.e. 
$S^{u}_{z,i} = S_{z,i}-S_{z,i}'$. Magnon contribution to
the dynamic structure factor is $ S_{u}(k,\epsilon) = 
S_u(k)\delta(\epsilon-\Omega_k)$, where  $S_{u}(k)= (u_{k}+v_{k})^{2}$.
At $J_{\perp}\gg J$ the structure factor is $S_u(k)\approx 1$.
The plot of $S_u(k)$ at $J_{\perp}=J$  is shown in Fig.2 (long-dashed line).

To find the interaction between the magnon and the phonon
we have to substitute transformed magnon operator 
$ t_{\alpha k} = u_{k} \tilde{t}_{\alpha k}
+  v_{k} \tilde{t}_{\alpha -k}^{\dagger}$
into the interaction term (g-term) of the Hamiltonian (\ref{H1}). This gives
\begin{eqnarray}
\label{hint}
H_{int}&=&\sum_{k,q,\alpha}\left\{
\Gamma_{k,q}\tilde t^{\dag}_{\alpha,k+q}\tilde t_{\alpha,k}\right.\nonumber\\
&+&\left. {1\over 2}\gamma_{k,q}(\tilde t^{\dag}_{\alpha,k+q}\tilde t^{\dag}_{\alpha,-k}
+\tilde t_{\alpha,-(k+q)}\tilde t_{\alpha,k})\right\}b_q + h.c.\\
\Gamma_{k,q}&=&g(u_{k+q}u_k+v_{k+q}v_k),\nonumber\\
\gamma_{k,q}&=&g(u_{k+q}v_k+u_kv_{k+q}),\nonumber
\end{eqnarray}
where $b_q$ is the Fourier transform of $b_i$.

We consider weak coupling with phonons, $g^2 \ll J^2$. Therefore
the magnon self energy, given by the diagram in Fig.3a, is of the form
\begin{equation}
\label{se}
\Sigma(k,\epsilon)= 
\sum_q {{\Gamma_{k,q}^2}\over{\epsilon-\omega_0-\Omega_{k-q}}}.
\end{equation}
Let us consider the limit $J_{\perp}\gg J$.
In this case $\Omega_k$ is given by (\ref{e}) and therefore
the integration in (\ref{se}) can be performed analytically.
This gives the correction to the triplet dispersion
\begin{equation}
\label{cor}
\delta \Omega_k=\Sigma(k,\epsilon=\Omega_k)= -{{g^2}\over
{\sqrt{(\omega_0-J\cos k)^2-J^2}}}.
\end{equation}
This answer is satisfactory if $\omega_0 > 2J$. However for lower phonon energy
($\omega_0 < 2J$) equation (\ref{cor}) gives an infinite correction at 
$k=k_c$, where $\cos k_c=(\omega_0-J)/J$. The reason for this is clear:
the integrand in (\ref{se}) has a pole and therefore in spite of the
weak coupling the perturbation theory is not valid in the vicinity of $k_c$.
To improve the situation we have to use the exact Green's function
\begin{equation}
\label{gt}
\tilde{G}(k,\epsilon)={1\over{\epsilon-\Omega_k-\Sigma^{\prime}(k,\epsilon)}},
\end{equation}
where $\tilde{G}$ is defined as
$\tilde{G}(k,t)=-i\langle T[\tilde{t}_{k,\alpha}(t),\tilde{t}_{k,\alpha}^{\dag}(0)]
\rangle$ \cite{G}.
The dispersion $\epsilon_k$ is given by the pole of this Green's function.
\begin{equation}
\label{zero}
[\tilde{G}(k,\epsilon)]^{-1}=0
\end{equation}
Note that the self energy $\Sigma^{\prime}$ in (\ref{gt}) is different from 
$\Sigma$ given by(\ref{se}). The difference is very simple: to get
$\Sigma^{\prime}$ one has to use the same eq. (\ref{se}) but with
$\Omega_k$ replaced by $\epsilon_k$.
This replacement is equivalent to summation of rainbow diagrams, Fig.3b.
Simple analysis shows that the vertex corrections (Fig.3c) are negligible
at small $g$. In the limit $J_{\perp} \gg J$ eq. (\ref{zero}) 
has an analytical solution. Away from $k_c$ this solution reproduces the 
perturbation  theory result(\ref{cor}), however at $k=k_c$ it gives a
discontinuity in dispersion. At $k=k_c+0$ the dispersion is pushed down as
$\epsilon_k \approx \Omega_{k_c}-(g^2/\sqrt{2J})^{2/3}$. Immediately on the left 
from $k_c$ ($k=k_c-0$) the dispersion disappears because of the huge decay width.
However away from this point ($k=k_c-\Delta k$) the width drops down
and at $\Delta k \gtrsim  g^4/(2J^2J_{\perp}^2\sin k_c)$ the halfwidth become
smaller than the energy. It means that the magnon reappears as a relatively narrow 
state.
At $J_{\perp} \sim J$ the equation (\ref{zero}) can be solved only numerically.
As an example we present in Fig.1 (solid line) the plot of the magnon
dispersion at $J_{\perp}=J$, $\omega_0=0.5J$, $g^2/J^2=0.2$. 
In spite of the relatively weak coupling to phonons the dispersion is 
substantially different from that without phonons (dashed line). 

The magnon contribution to the
the dynamic structure factor is 
$S_u^{(m)}(k)\delta(\epsilon-\epsilon_k)$, where  $S_{u}^{(m)}(k)= Z_k
(u_{k}+v_{k})^{2}$, and
$Z_k=\left(1-{{\partial \Sigma^{\prime}}\over{\partial \epsilon}}\right)^{-1}$ is
the quasiparticle residue. The plot of $S_u^{(m)}(k)$ is presented in Fig.2 by solid
line. It is also very much different from that without phonons (long-dashed line).
Concluding  the discussion of magnons we would like to note that 
the discontinuity in the spectrum is similar to the roton spectrum endpoint in liquid 
$^4$He, see e.g. Ref. \cite{Lifshitz}.

Consider now a  phonon-magnon bound state.
The binding arises due to the diagrams shown in Figs.4a,b.
To solve the problem in general case one has to use the Bethe-Salpeter equation
with account of retardation. However for small phonon-magnon coupling ($g^2 \ll J^2$)
which we consider in the present work, the binding is relatively weak and we can use 
simple second order Schrodinger perturbation theory. In this approximation the 
diagrams Fig.4a,b give the following effective interaction
\begin{eqnarray}
\label{M}
&&M(Q,k,p)=M_a(Q,k,p)+M_b(Q,k,p)=\\
&&{{\Gamma_{k,Q-p}\Gamma_{p,Q-k}}\over{E_Q-2\omega_0-\epsilon_{k+p-Q}}}+
{{\gamma_{k,Q-p}\gamma_{p,Q-k}}\over{E_Q-\epsilon_k-\epsilon_p-\epsilon_{k+p-Q}}},
\nonumber
\end{eqnarray}
where $E_Q$ and $Q$  is the energy and the momentum of the bound state. As soon as the
interaction is known the bound state is defined by usual equation
\begin{equation}
\label{BS}
\left[E_Q-\omega_0-\epsilon_k\right]\psi(k)=
\int {{dp}\over{2\pi}}M(Q,k,p)\psi(p).
\end{equation}
In the limit $J_{\perp}\gg J$ the kernel is very simple 
$M \approx M_a \approx -g^2/(\omega_0+2J\cos^2{Q/2})$, and
solution of equation (\ref{BS}) gives $E_Q=J_{\perp}-J+\omega_0-\epsilon_b$,
with the wave function and binding energy equal to
\begin{eqnarray}
\label{eb}
\psi(p)&=&{{\left(8J\epsilon_b^3\right)^{1/4}}\over{\epsilon_b+2J\cos^2{p/2}}}\\
\epsilon_b(Q)&=&{{g^4}\over{2 J (\omega_0+2J\cos^2{Q/2})^2}}.\nonumber
\end{eqnarray}
At $J_{\perp} \sim J$ equation (\ref{BS}) can be solved only numerically.
We present it Fig.1 (dashed line) the plot of the bound state 
dispersion at $J_{\perp}=J$, $\omega_0=0.5J$, $g^2/J^2=0.2$. Shadowed areas
show phonon-magnon continuum. It is clear that binding is maximum at $Q=\pi$.

The bound state has spin $S=1$ and therefore it can be excited in neutron 
scattering. The mechanism for this excitation is shown in Fig.4c: first a
virtual magnon is excited and then it emits a phonon with formation
of the bound state. The corresponding contribution to the dynamic structure factor is
$S_u^{(b)}(k)\delta(\epsilon-E_k)$ with $S_u^{(b)}(k)$ given by
\begin{equation}
\label{suu}
 S_{u}^{(b)}(k) = \left|{{(u_k+v_k)}\over{E_k-\Omega_k-\Sigma^{\prime}(k,E_k)}}
\sum_{p} \Gamma_{k,k-p}\psi(p) \right|^{2}.
\end{equation}
At $J_{\perp} \gg J$ integration in this formula with account of eqs. (\ref{eb}) gives
\begin{equation}
 S_{u}^{(b)}(k) = {{g^2\sqrt{2\epsilon_b(k)/J}}\over{(\omega_0-2J\cos^2{k/2})^2}}.
\end{equation}
It is assumed that the denominator in this formula is not too small. At small
denominator there is strong mixing between the magnon and the bound
state and therefore a more accurate consideration is necessary. 
The structure factor $ S_{u}^{(b)}(k)$ for $J_{\perp}=J$, $\omega_0=0.5J$ and
$g^2=0.2J^2$ obtained by numerical integration in (\ref{suu}) is shown in Fig.2
by the dashed line. At $k=\pi$ the bound state excitation probability is 10
times smaller than that for the magnon. However at $k\sim 0.8\pi$ the probabilities
are comparable.

Above we have considered the magnon and the bound state of the magnon
with a phonon. In addition to these states there are bound states of two
magnons considered in Ref. \cite{us1}. The dynamic phonon does not qualitatively
influence these bound states. However additional phonon diagrams shown in
Fig.5  push down the energies of these states. To illustrate this 
for $J_{\perp}=J$, $\omega=0.5J$ and $g^2=0.2J^2$ we
show in Fig.1 (dashed-dotted line) the dispersion of the two-magnon bound state with 
$S=1$. Because of the decay to the phonon and two magnons this
state attains finite width $\Gamma \sim 0.15J$ \cite{ug}. This is why the 
state disappears at $k\sim 0.6\pi$ where splitting from the corresponding continuum is
getting smaller than halfwidth. The structure factor for this state is
$S_g(k,\epsilon)=S_g(k)\delta_{\Gamma}(\epsilon-E_k)$, where $\delta_{\Gamma}$
is a function with finite width. The index $g$ indicates that the state
can be excited at $k_{\perp}=0$, i.e. by ${\bf S_{i}+S^{\prime}_{i}}$. The formula
for $S_g(k)$ is presented in \cite{us1}. Integration in this formula  gives the
plot shown in Fig.2 by the dotted-dashed line. At $k=\pi$ the excitation probability
is 4 times smaller than that for the magnon. However at 
$k \sim 0.8\pi$ they become equal.

In conclusion, we have studied the two-leg
antiferromagnetic Heisenberg ladder with spins coupled to
dynamic phonons. In the analysis we use the bond operator representation
together with the diagram technique and Brueckner approximation.
It has been demonstrated that the  phonons
qualitatively change spectrum of triplet excitations:
1) An additional triplet appears as a  magnon-phonon bound state. 
Dispersion of this triplet for some particular parameters is shown
in Fig.1 (dashed line); 
2) the  magnon (elementary triplet) dispersion attains a discontinuity, Fig.1, 
solid line; 
3)k-dependence of the 
spin structure factor (Fig.2, solid line) is substantially different from
that without phonons (Fig.2, long-dashed line).

I would like to thank M. Kuchiev for stimulating
discussions.

\begin{figure}
\caption
{Triplet
excitations on the isotropic ($J_{\perp}=J$)
ladder. The long-dashed line is the magnon dispersion obtained by
the Brueckner method without coupling to phonons. Dots represent the
same dispersion obtained in Ref.[5] using series expansion.
Solid line gives the magnon dispersion with account of
coupling to phonons. The phonon energy is $\omega_0=0.5J$ and the
coupling constant $g^2/J^2=0.2$. The dashed line corresponds to
the bound state of the magnon and the phonon. The
dashed-dotted line corresponds to the bound state of the two magnons
with total spin S=1.
Shadowed areas gives magnon-phonon continuum.}
\label{fig.1}
\end{figure}

\begin{figure}
\caption
{Spin structure factors (excitation probabilities) for isotropic ladder, 
$J_{\perp}=J$.
The long-dashed line corresponds to the magnon without
coupling to phonons. Solid line describes the same magnon
coupled to phonons ($\omega_0=0.5J$, $g^2/J^2=0.2$). The dashed
line describes bound state of the magnon and the phonon.
The dashed-dotted line corresponds to the bound state of the two magnons.}
\label{fig.2}
\end{figure}
 
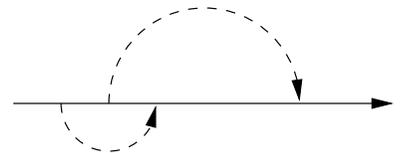
\begin{figure}
\caption
{Magnon self energy due to the phonons. Dashed line corresponds to the
phonon Green's function. (a) Single  loop,
(b) rainbow diagrams, (c) vertex correction.}
\label{fig.3}
\end{figure}

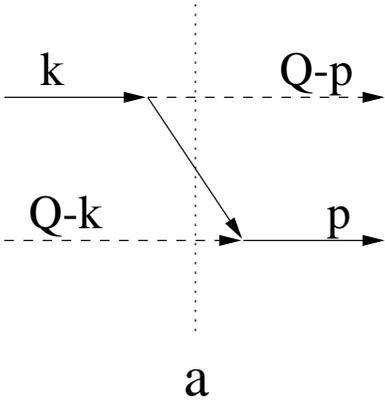
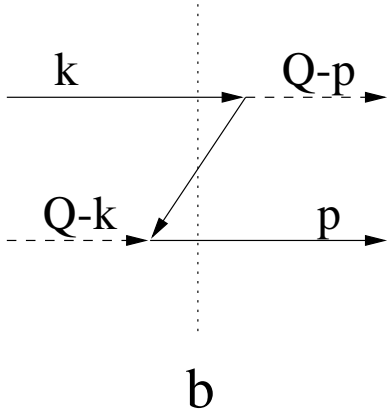
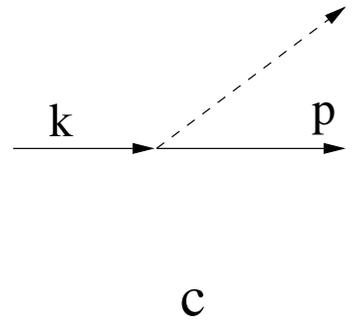
\begin{figure}
\caption
{Diagrams (a) and (b) represent the phonon - magnon scattering amplitudes
responsible for the binding. Solid line describes the magnon and dashed line
describes the phonon.
The diagram (c) shows mechanism for
 the bound state contribution to the spin structure factor.}
\label{fig.4}
\end{figure}

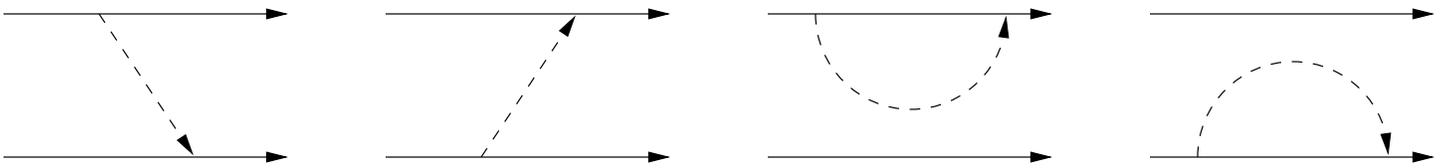
\begin{figure}
\caption
{Phonon corrections for the two magnon bound state.}
\label{fig.5}
\end{figure}


\begin{references}

\bibitem{Dagotto} For a review see 
E. Dagotto and T.M. Rice, Science {\bf 271}, 618 (1996).

\bibitem{Copalan} S. Gopalan, T.M. Rice, and M. Sigrist,
 Phys. Rev. B {\bf 49}, 8901 (1994).

\bibitem{Exact}T. Barnes and J. Riera, Phys. Rev. B {\bf 50}, 
6817 (1994).

\bibitem{DMRG} S.R. White, R.M. Noack, and D.J. Scalapino,
 Phys. Rev. Lett. {\bf 73}, 886 (1994).
 
\bibitem{O} J. Oitmaa, R.R.P. Singh, Z. Weihong. Phys. Rev. B
{\bf 54}, 1009 (1996).

\bibitem{Garrett} A. W. Garrett {\it et al.}, Phys. Rev. B {\bf 55},
3631 (1997); Phys. Rev. Lett. {\bf 79}, 745 (1997).

\bibitem{BS} 
G. Bouzerar, A.P. Kampf, and G.I. Japaridze, Phys. Rev. B {\bf 58}, 3117 (1998);
 T. Barnes, J. Riera, and D.A. Tennant, cond-mat/9801224.

\bibitem{Dam} K. Damle and S. Sachdev, Phys. Rev. B {\bf 57}, 8307 (1998).

\bibitem{us1} O. P. Sushkov and V. N. Kotov, Phys. Rev. Lett. {\bf 81},
1941 (1998).

\bibitem{us2} V.N. Kotov, J. Oitmaa, O. P. Sushkov, Zheng Weihong, 
cond-mat/9903154.

\bibitem{Affleck} I. Affleck, in"Dynamical Properties of Unconventional Magnetic 
Systems" [ed. A. Skjeltorp and D. Sherrington, Kluwer Academic Publishers, 
Dordrecht, 1998] p. 123; see also cond-mat/9705127.

\bibitem{Uhrig} G. S. Uhrig, F. S. Sch\"{o}nfeld, M. Laukamp and E. Dagotto, 
Europhys. J. B {\bf 7}, 67 (1999).

\bibitem{Cross} M. C. Cross and D. S. Fisher, Phys. Rev. B {\bf 19},
402 (1979).

\bibitem{Kh} D. Khomskii {\it et al}., Czech. J. Phys. {\bf 46}, Suppl. S6, 32
(1996).

\bibitem{Wel} G. Wellein, H. Fehske, and A. P. Kampf,  Phys. Rev. Lett. 
{\bf 81}, 3956 (1998).

\bibitem{K} V.N. Kotov, O. Sushkov, Z. Weihong, and J. Oitmaa,
Phys. Rev. Lett., {\bf 80}, 5790 (1998).

\bibitem{G} Note that the Green's function (\ref{GN}) is defined without
tilde: $G(k,t)=-i\langle T[t_{k,\alpha}(t),t_{k,\alpha}^{\dag}(0)]\rangle$.

\bibitem{Lifshitz} E. M. Lifshitz and L. P. Pitaevskii, {\it Statistical Physics}
(L. D. Landau and E. M. Lifshitz Course of Theoretical Physics, Vol. 9),
Pergamon Press, Oxford, 1986.

\bibitem{ug} This bound state can not decay into phonon and single magnon
because it has positive parity with respect to the ladder leg
permutation.


\end{references}
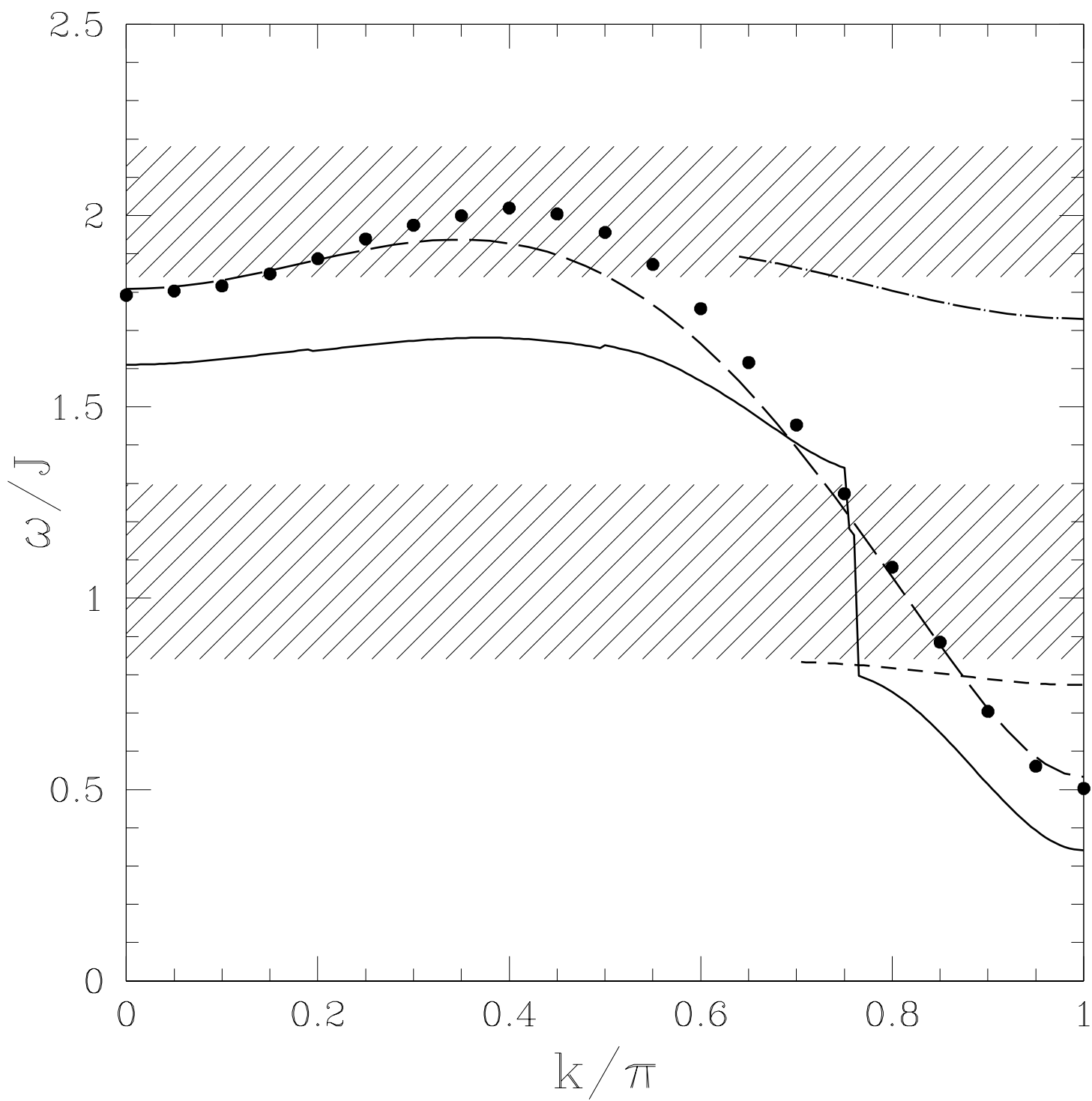
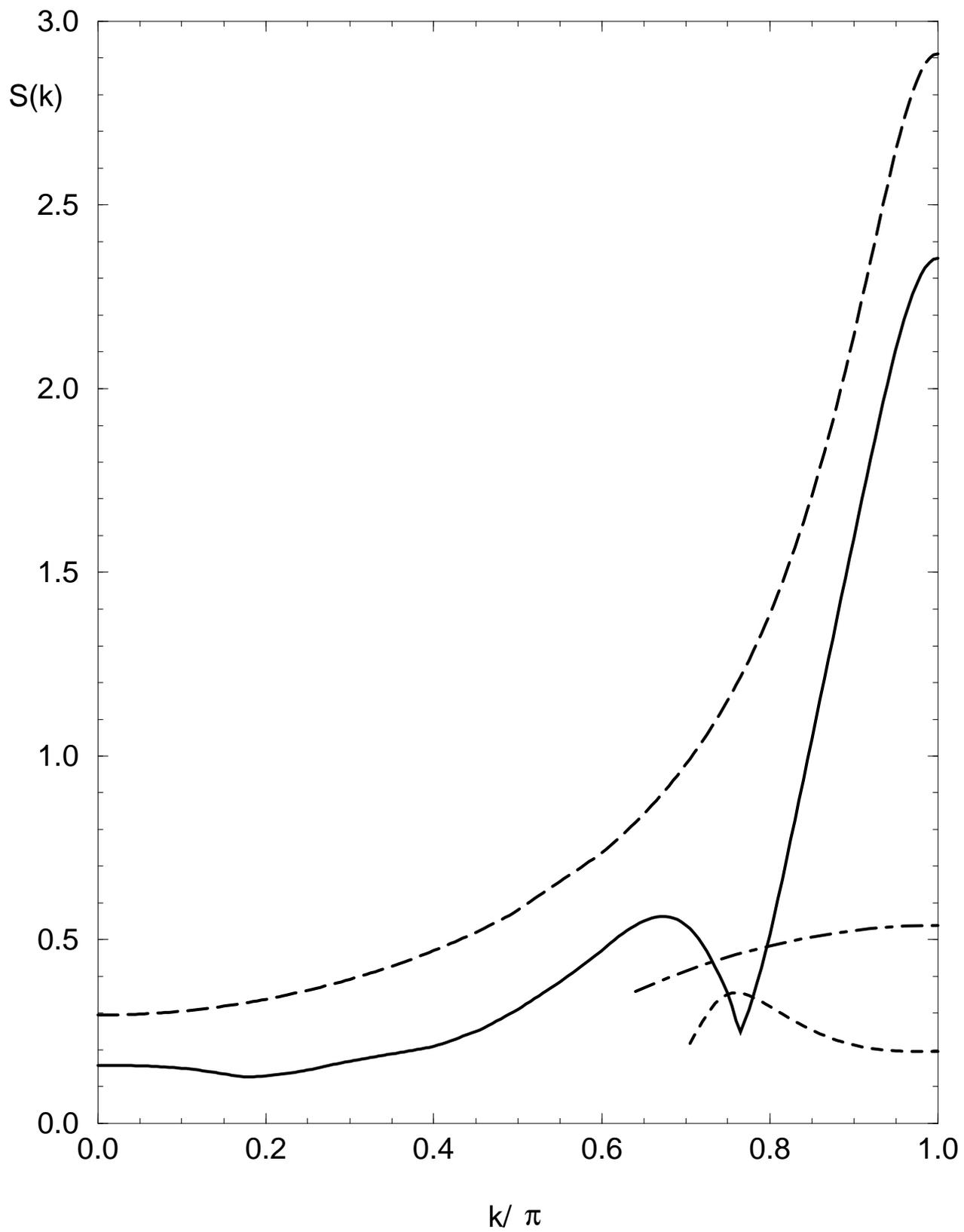
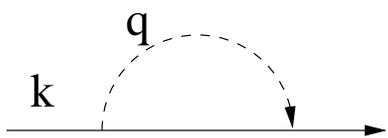
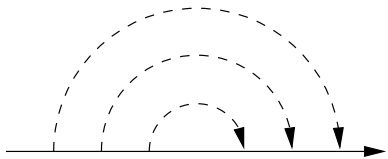
\end{document}